\documentclass[preprint2]{emulateapj}

\shorttitle{SAOLIM, a prototype of a low cost System for Adaptive
Optics with Lucky Imaging. Design and performance.}
\shortauthors{aceituno et al.}

\begin{document}


\title{SAOLIM, a prototype of a low cost System for Adaptive
Optics with Lucky Imaging.\\ Design and performance.}


\author{J. Aceituno and S. F. Sanchez}
\affil{Centro Astronomico Hispano-Aleman (CAHA), Jesus Durban
Remon 2-2, 04004 Almeria, Spain}

\author{J. L. Ortiz and F. J. Aceituno}
\affil{Insituto de Astrofisica de Andalucia, CSIC Granada, Spain}
\email{aceitun@caha.es}



\begin{abstract}
A prototype of a low cost Adaptive Optics (AO) system has been
developed at the Instituto de Astrofisica de Andalucia (CSIC) and
tested at the 2.2m telescope of the Calar Alto observatory. We
present here the status of the project, which includes the image
stabilization system and compensation of high order wavefront
aberrations with a membrane deformable mirror. The image
stabilization system consists of \textbf{magnet} driven tip-tilt
mirror. The higher order compensation system \textbf{comprises of}
a Shack-Hartmann sensor, a membrane deformable mirror with 39
actuators and the control computer that allows operations up to
420Hz in closed loop mode. We have successfully closed the high
order AO loop on natural guide stars. An improvement of 4 times in
terms of FWHM was achieved. The description and the results
obtained on the sky are presented in this paper.
\end{abstract}


\keywords{Adaptive optics,micromachined membrane deformable
mirror, EMCCD, low cost.}



\section{Introduction}

The use of adaptive optics in astronomy with reasonable spatial
order correction and temporal bandwidths has been restricted for
many years to observatories with a high instrumentation budget. In
current versions, such systems use expensive deformable mirrors,
and digital signal processors to apply the reconstruction
algorithms. This results in  complex systems
\citep[e.g.][]{hip2000}, taking several years to be developed.
Nowadays, the total cost can be reduced by several orders of
magnitude \cite[e.g.][]{dai1999} thanks to the availability of
relatively cheap membrane deformable mirrors \cite{vdo1995},
single photon detectors with a \textbf{reasonably} high frame rate
based on the EMCCD technology \citep[e.g.][]{dus2004} and low cost
Tip-Tilt systems.

In this paper we describe a System of Adaptive Optics with Lucky
Imaging (hereafter SAOLIM). We present the design, construction
and results of a low order AO system for 1-2m class telescopes,
almost entirely developed with available commercial components,
and with a total cost of $\sim$ 35000 euros in hardware
components.

The optical design enables a FOV of 90x90 arcsec$^{2}$ for the
scientific camera. SAOLIM is optically corrected and transparent
for a wavelength range between 1.0-2.5$\mu$m.

Our system is based on a membrane deformable mirror
\citep[e.g.][]{pat2000}, and a single PC to perform all the
computations and hardware control, integrating everything in a
simple and compact design as we will see later. The dual
wireless/ethernet communication of the device allows an easy
setup, because no cabling has to be installed at the telescope,
reducing a considerable amount of possible problems.

This instrument can be used as an input correction for
applications where reaching the diffraction limit of the telescope
is required, such as a lucky imaging system, like ASTRALUX
\cite{hor2008}, or as a complement for the shift-and-add approach
\citep[e.g.][]{bates1980}. This device can take advantage of a low
order AO system, in such a way that the rate of the useful images
could be increased, and therefore the performance of the
instrument can be improved. Such innovative technique has been
tested recently at the Palomar observatory, getting excellent
results \citep[e.g.][]{law2008}.

\begin{table*}
\begin{center}
\caption{Main Instrument Parameters\label{tbl-2}}
\begin{tabular}{ll}
\tableline\tableline AO SYSTEM \\
\tableline\tableline focal ratio: & cassegrain f:8 \\
principle of operation: & Adaptive optics system \\
AO closed-loop sample speed: & 420Hz\\
modes of reconstruction: & 15\\
 & \\
WAVEFRONT SENSOR \\
\tableline\tableline principle of operation: & Shack-Hartmann sensor \\
detector: & EMCCD Andor IXON DU860 \\
Chip size: & 128x128 pixels\\
lenslet-arrays: & 5x5{\tablenotemark{a}} and KS28{\tablenotemark{b}}\\
pixel scale: & 0.45 arcsec/pixel\\
\textbf{FOV of each spot:} & 10.5 arcsec\\
wavelength range: & 400-950 nm \\
filter wheel: & 6 positions. Neutral density filters \\
 & \\
DEFORMABLE MIRROR \\
\tableline\tableline principle of operation: & membrane deformable mirror \\
actuators: & 39\\
diameter: & 30mm\\
maximun depth: & 8$\mu$m{\tablenotemark{c}}\\
voltage range: & 0-250V\\
sample speed: & 1kHz\\
reference voltage: & 180V\\
Useful diameter: & 20mm{\tablenotemark{c}}\\
 & \\
TIP-TILT SYSTEM \\
\tableline\tableline principle of operation: magnetic pivots\\
model: & AO-7 SBIG\\
sample speed: & 50Hz\\
 & \\
SCIENTIFIC CAMERA \\
\tableline\tableline principle of operation: & back-iluminated CCD 1024x1024 \\
scientific camera FOV: & 90x90 arcsec$^{2}$ \\
pixel scale: & 0.08 arcsec/pixel\\
pixel size: & 24$\mu$m\\
filter wheel: & 4 positions 50mm dia. each\\
wavelength range: & 950-2500 nm \\
 & \\
 CONTROL \\
 \tableline\tableline principle of operation: & Single PC INTEL
 pentium IV 3.4Ghz, 2Gb RAM \\
\tableline\tableline

\end{tabular}
\tablenotetext{a}{5x5 hexagonal microlenses.}
\tablenotetext{b}{keystone-shaped microlenses.}
\tablenotetext{c}{\textbf{According to the manufacturer.}}
\end{center}
\end{table*}
\begin{figure*}
\includegraphics[scale=.70, clip=true]{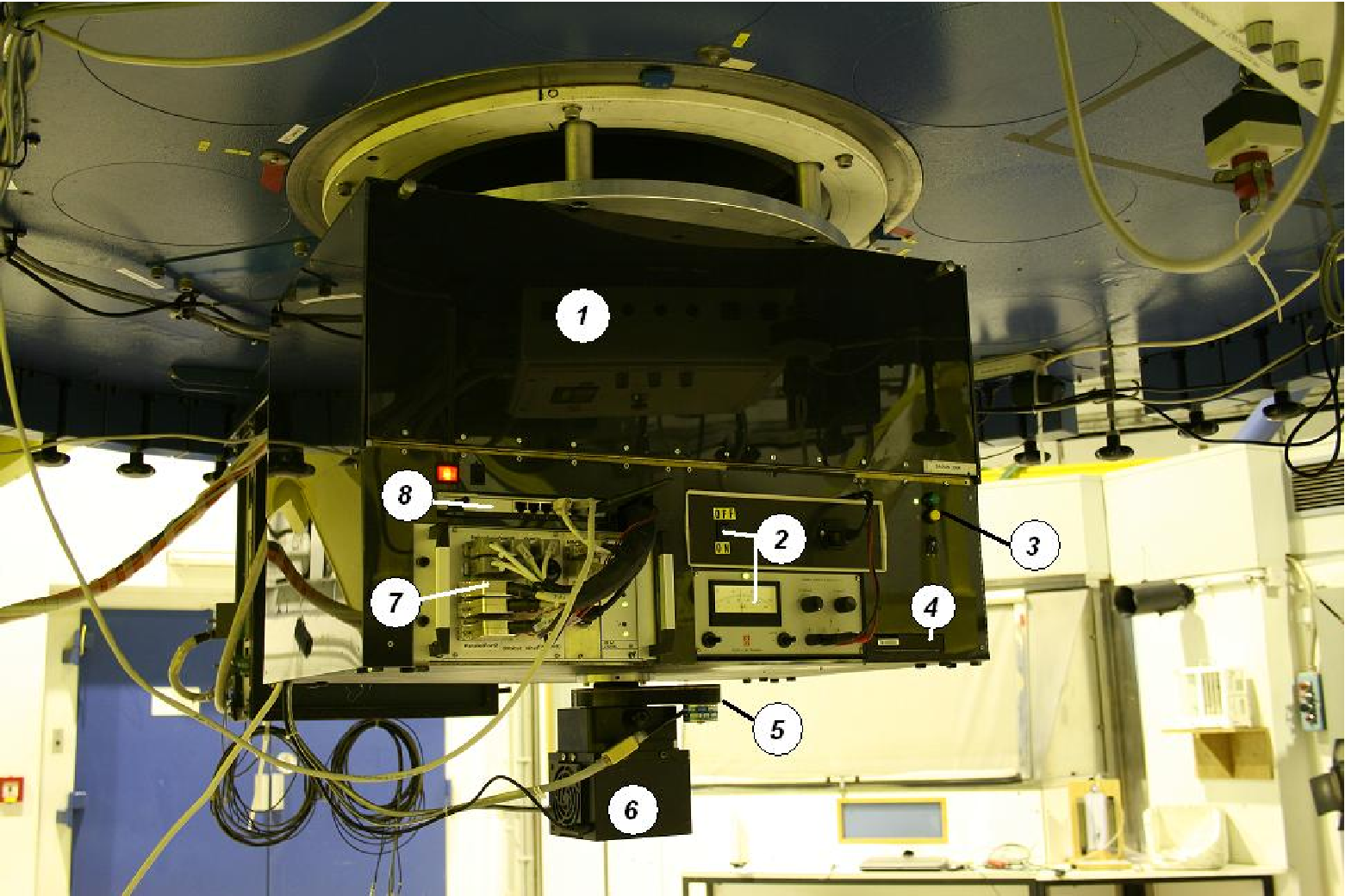}
\caption{SAOLIM mounted at the Calar Alto 2.2m Telescope
Cassegrain Focus. Different elements are labelled: (1) Optical
bench, (2) DM electronics, (3) PC Pentium IV, (4) hard disk, (5)
scientific filter wheel, (6) scientific camera, (7) hardware
electronic, (8) router. The instrument measures 75x40x50cm and
weights 70kg.}
\includegraphics[scale=.70, clip=true]{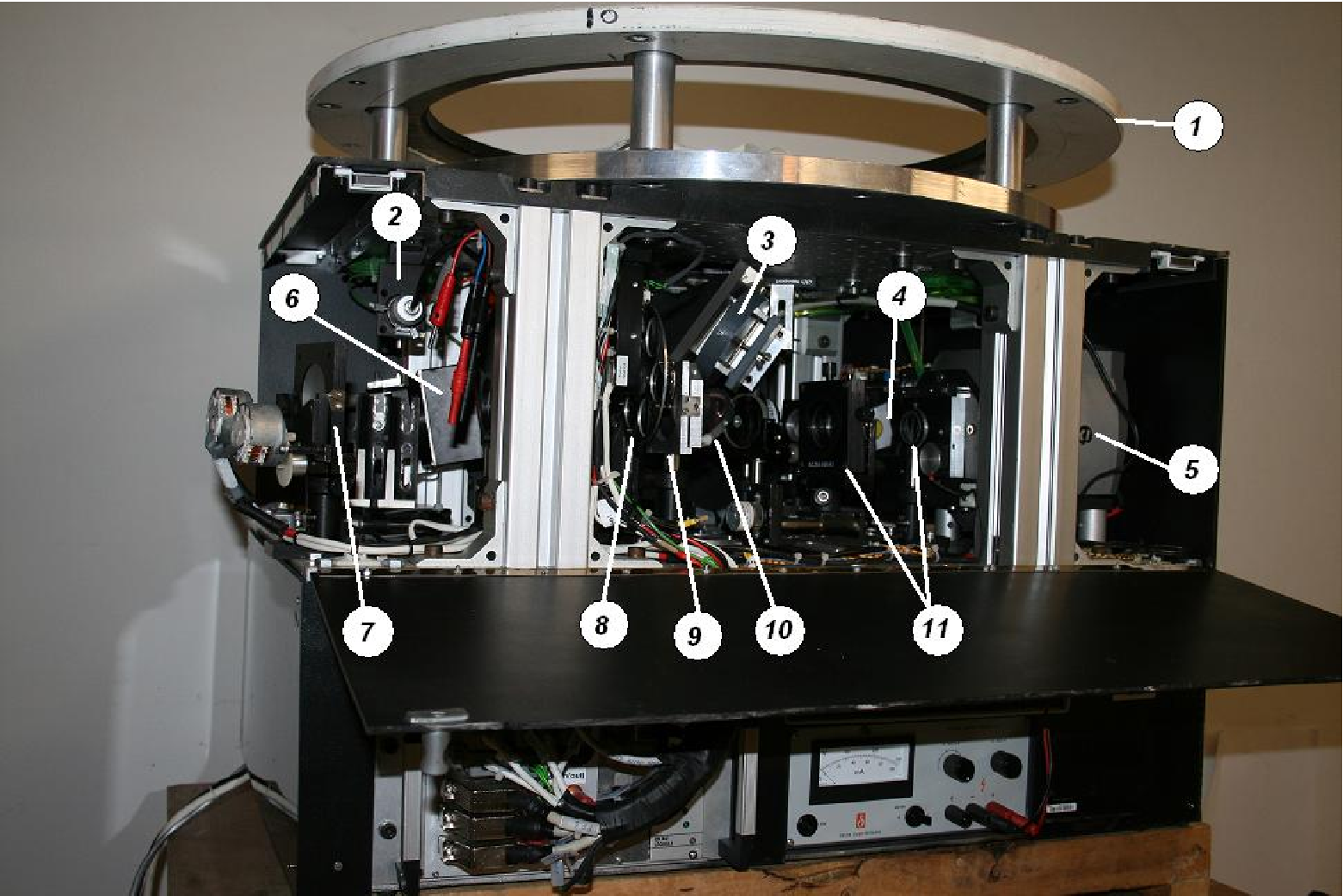}
\caption{Lateral picture of the optical bench in SAOLIM: (1)
Cassegrain adapter, (2) \textbf{a reference fiber located at f/8
focus}, (3) fold mirror, (4) membrane deformable mirror, (5)
wavefront sensor, (6) Tip-Tilt mirror, (7) fold mirror, (8)
wavefront sensor filter wheel, (9) lenslet array, (10) beam
splitter, (11) wavefront sensor relay lens. }

\end{figure*}

\section{System description}

Figure 1 shows a picture of the instrument attached to the
telescope. Labels indicating the main components have been
included. A detail of the optical components is shown in Figure 2
where a view of the inside of the instrument (located at the lab)
is presented. The main optical elements have been labelled in this
figure.
\subsection{Mechanical and optical design.}

A sketch of the optical setup is shown in Figure 3. The optical
design is similar to that of ALFA, the AO system that was operated
at the 3.5m telescope of the Calar Alto observatory since 1997
until 2005 \citep[e.g.][]{hip2000}. A Shack-Hartmann wavefront
sensor (hereafter SHS) is placed optically conjugated with the
membrane deformable mirror (MDMM) and with the entrance pupil of
the whole system. \textbf{To align the pupil on the DM, a camera
is temporarily located in different parts of the optical axis,
until a sharp image of the main mirror is obtained.}Two achromatic
doublets (E1) separated by 5mm, with focal distances of 300mm each
one conjugate the entrance pupil over a 20mm diameter circular
area on the deformable mirror. Another pair of achromatic doublet
lenses (E3 and E6) is configured as a Kepler telescope and
conjugates the membrane's selected zone on the SHS. The SHS
consists of a microlenses array (E7, focal length 45mm) and two
achromatic doublet lenses (E8 and E9) to re-image the subapertures
spots on the EMCCD with an appropriate pixel scale yielding a
value of 0.48"/pixel. An aplanatic lens (E2) is placed between the
MMDM and E3 to reduce coma aberration in the Kepler telescope. A
dichroic beam-splitter (E4) reflects the near infrared part of the
spectrum to the scientific camera whereas the visible part to go
to the SHS. Finally, a motorized flat mirror (E5) folds the
optical path to keep the design compact.

The optical performance of the design was evaluated with the
optical software ZEMAX \footnote{http://www.zemax.com}. Figures 4
and 5 show the spots sizes at different field angles at the
scientific camera of SAOLIM. The geometry and sizes of the spots
are comparable to those of the Airy disk in different angles,
whose amplitudes have been selected to match the isoplanatic angle
in the K-Band. The Strehl ratio expected is about 72\%, close to
the theoretical value of 82\%, foreseen by ZEMAX. The image
quality is almost constant through the different angles, with a
wavefront distortion smaller than $\frac{\lambda}{4}$ (Fig 4). The
system is equipped with an artificial point source fed by an
optical fiber and has other movable motorized components such a
filter wheel, shutters, and focuser of the wavefront sensor (WFS,
hereafter).

\textbf{Finally, the Tip-Tilt mirror is located just before the
DM. The only drawback of not being located in the pupil plane is
that it is moved around the ground-layer turbulence. According to
ZEMAX this produces a shift in the science camera's pupil of less
than 1\% of its size, for a typical atmospherical Tip-Tilt. This
wouldn't harm even coronagraphic observations with an undersize
stop. On the positive side, at least two optical elements (which
would be required to re-image the pupil) are saved. Similar design
have been adopted in other working AO systems
\citep[e.g.][]{hip2000}, \citep[e.g.][]{pet2010}}

\begin{figure}
\includegraphics[scale=.35, clip=true]{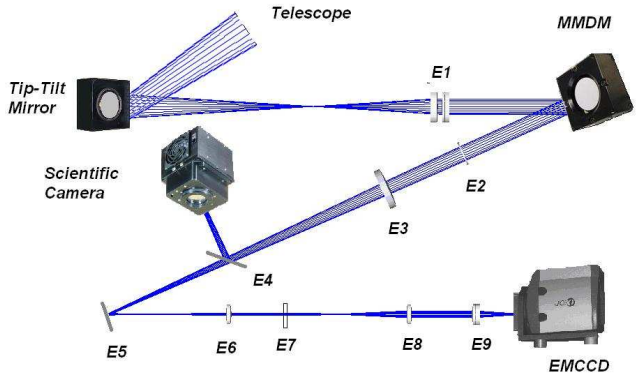}
\caption{The Figure shows the optical design of SAOLIM.}

\end{figure}
\begin{figure}
\includegraphics[scale=.55, clip=true]{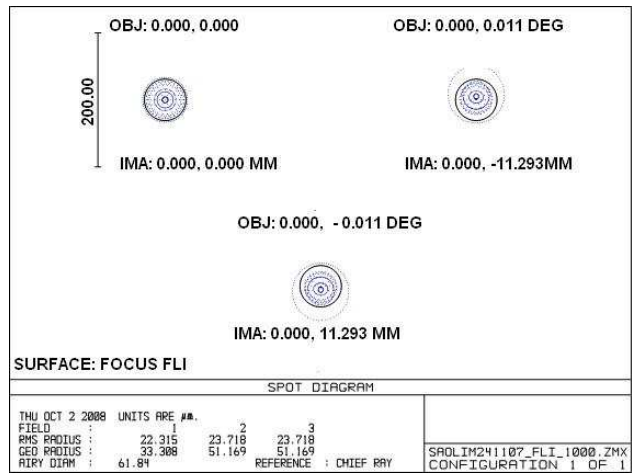}
\caption{Spot geometry at the scientific camera at different entry
angles. \textbf{The field of view is 90 arcseconds.} Values
estimated by Zemax.}

\end{figure}




\subsection{Wavefront sensor.} \label{bozomath}


The wavefront reflected by the deformable mirror can be sampled by
two different lenslet arrays with different configurations. The
first one is a keystone-shaped array with 28 microlenses (Figure
6) which is detected by a 128x128 pixels EMCCD camera. The focal
length of the microlenses is 45mm. This geometry allows for an
optimum filling of the annular telescope aperture contrary to
other designs, like an hexagonal or square grid. In addition, if
the subapertures in the different rings are designed in such a way
that all of them have the same area, the spots are equally bright,
and the noise pattern is uniform. The reconstruction benefits of
this configuration with respect to other more usual ones (e.g.
hexagonal-shaped) are described by \cite{kas2003}.

A second array comprises a 5x5 hexagonal-shaped lens configuration
(Figure 6) with a focal length of 36.1mm and 1mm pitch. The focal
length is different than in the previous case, so a linear
motorized stage can place the relay lenses in such a way that the
pixel scale remains constant at the detector. This setup can be
used with fainter targets due to the smaller number of microlenses
than the previous one.

A variety of wave-front reconstruction algorithms using data
provided by SHS are available \cite{lij2002}. The modal
reconstruction algorithm will be better in case of low
signal-to-noise ratio (SNR) conditions \cite{lij2002}. For that
reason we used the modal one for our project. In this procedure
the measured focal position of each microlens is used to determine
the local wave-front gradients, in such a way that the wavefront
shape can be reconstructed by means of a vector of coefficients in
a polynomial basis. The Karhunen-Loeve polynomials are used here.

The modal reconstruction algorithm is described in \cite{sou1980}.
Here we summarize the main steps of this procedure. The desired
coefficient vector \emph{a}, representing the reconstructed
wave-front, can be derived from the array of measured wavefront
gradients, by applying an inversion method:

\begin{equation}
 a={\bf (A^\dagger\cdot A)^{-1}A^\dagger\cdot S},
\end{equation}

where \emph{A} is a rectangular matrix with \emph{2N} rows and
\emph{M} columns (with \emph{N} being the number of microlens and
\emph{M} the total number of Karhunen-Loeve terms used). Its
coefficients can be calculated from the partial derivatives of the
polynomials \cite{dai1995}. The local slopes can be organized to
form a slope vector \emph{S} of size \emph{2N} \cite{sou1980}.

\subsection{The Deformable and Tip-Tilt mirrors.} \label{bozomath}




The micro-machined membrane deformable mirror consists of a chip
with a silicon nitride membrane coated with aluminium. It was
manufactured by OKO Technologies \cite{vdo1995}. The membrane
shape is driven electrostatically by the voltages applied to 39
control electrodes. Since the force between the membrane and the
electrodes is attractive, the membrane can be pulled only toward
its base. Therefore deformations in both directions can be led by
biasing the mirror to a nonzero voltage. The reference value is
180 V for this bias, with an effective mirror diameter of 20mm
according to the manufacturer.

The membrane is coated with an evaporated layer of aluminium to
make it reflective and conductive. Two digital drivers provide an
8-bit voltage control for the output channels, whereas two
high-voltage boards amplify the digital signals (0-250V) that are
subsequently applied to each electrode.
\begin{figure}
\includegraphics[scale=.60, clip=true]{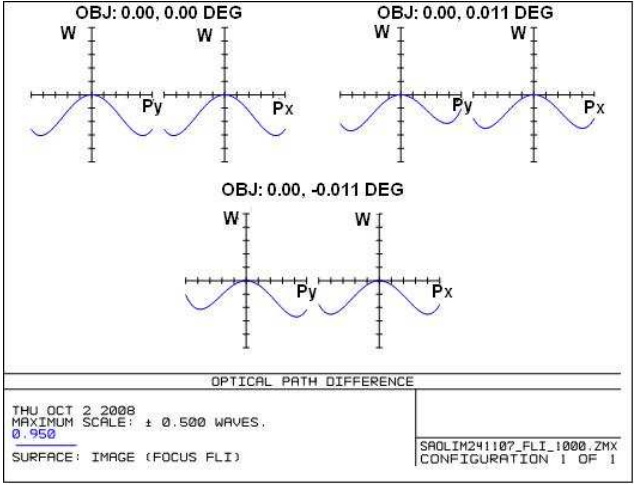}
\caption{A ray fan diagram showing the aberrations of different
angles \textbf{($\pm39.6$ arcseconds)} at the scientific camera of
SAOLIM, estimated by Zemax.}

\end{figure}
\begin{figure}
\includegraphics[scale=.40, clip=true]{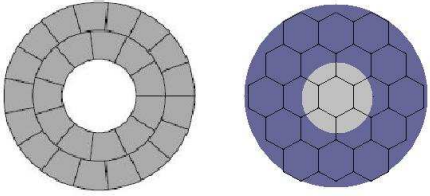}
\caption{Left panel: Keystone-shaped lenslet array in 28 microlens
configuration. Right panel: 5x5 lenslet array configuration. The
blue area depicts the telescope pupil. \label{fig2}}
\end{figure}

The MDMM control is based on the previous knowledge of the so
called influence functions. These functions are the responses of
the membrane to the action of one particular isolated electrode.
They were obtained by direct measurements of the wavefront, using
the SHS system, when the highest voltage is applied to each
actuator keeping the rest to the bias level \cite{cla1986}. The
surface's wavefronts are expressed as a Karhunen-Loeve's
polynomial expansion with 14 terms (including the Tip-Tilt), which
corresponds to the fourth order of these polynomials. The set of
functions can be grouped into the so-called influence functions
matrix (\emph{IFM}). Assuming that the total deflection of the
mirror is a linear superposition of the deflections resulting from
each control channel, we can obtain the shape of the membrane as a
response to a given set of voltages applied in the electrodes by a
simple matrix multiplication:

\begin{equation}
 IFM\ast V=a ,
\end{equation}

where vector \emph{V} is the set of \emph{k} squared voltages
applied to each electrode and \emph{a} is the shape of the
membrane expressed in terms of an expansion of Karhunen-Loeve's
polynomials \cite{dai1995}:

The \emph{control matrix} \emph{CM} is obtained as the inverse
matrix of \emph{IFM}. It relates the vector of coefficients (S) to
the required voltages by:

\begin{equation}
 CM*S=V .
\end{equation}

Because of the particular shape of the mirror, the \emph{CM} is
not a square matrix. Therefore a pseudo-inversion procedure to the
\emph{IFM} matrix is required in order to derive the \emph{CM}.
The Singular value decomposition method (\emph{SVD}) was adopted.
Some membrane modes may be removed by setting to zero a singular
value in the \emph{IFM} during the inversion process to avoid
infinite values. This operation reduces the capability to
reproduce some surfaces but it makes the control of the mirror
more stable.

Finally, the procedure to create different surfaces or to
compensate for the effects of the turbulence is applied
iteratively. \textbf{The iterative process is a negative feedback
loop and is similar to the one used by other AO systems
\citep[e.g.][]{hip2000}.} This allows us to obtain a better
performance than with a single iteration, due to the non-linearity
effects. Therefore the wavefront sensing errors or any
overshooting in the applied voltages are minimized during this
iterative operation (Closed-loop).

The set of 39 voltages \emph{$V_n$} \textbf{at instant \emph{n}}
is given by

\begin{equation}
  V_n=V_{n-1}+ \alpha \cdot \omega(CM \cdot {\phi})\\,
\end{equation}

where \emph{$\phi$} contains the SHS measurements of the wavefront
shape expressed in terms of an expansion of the Karhunen-Loeve's
polynomials (\emph{$F_{i}(x,y)$}) \emph{w} is a vector of weights
for each mode, and \emph{$\alpha$} is a damped parameter with
values between 0 and 1. The finally adopted value for
\emph{$\alpha$} was 0.75, derived empirically to grant the
convergence of the iterative process. The surface generated with
this set of voltages is the closest solution, in the least-squares
sense, to the surface \emph{S}. The root mean square (\emph{RMS})
of the residual can be expressed by:

\begin{equation}
RMS = |\sum_{i}\phi_{i}F_{i}(x,y)-\sum_{i}\phi'_{i}F_{i}(x,y)|\\.
\end{equation}

Therefore,

\begin{equation}
RMS=|\sum_{i}(\phi_{i}-\phi'_{i})F_{i}(x,y)|=\sqrt{(\phi_i-\phi'_i)^2}.
\end{equation}

The Tip-Tilt can be compensated with the deformable mirror, at the
expense of consuming a substantial fraction of the dynamic range
of the actuators. As an alternative solution, a steering mirror is
used for that purpose.

The Tip-Tilt mirror was manufactured by SBIG
\footnote{http://www.sbig.com}. It consists of a steering mirror
capable of achieving motion rates up to 50Hz. This Tip-Tilt mirror
has magnets on the back size. That interact with the current
flowing through a set of voice coils on the housing module to
rapidly move the mirror. The technique is very similar to that
employed in loudspeakers, except that there the magnet is fixed
and the wires are on the speaker cone. The mirror and magnets are
suspended using a flexible beryllium copper membrane. A needle
pushes up against a jewel bearing mounted at the center of the
mirror to hold the focus constant. SBIG has developed a
proprietary technique to rapidly damp the motion of the mirror, so
small movements are precise, with very little overshoot or
fluctuations. The tilt of the mirror during operation is very
small, and it does not lead to any measurable de-focus at the
edges of the frame. The correction range of the Tip-Tilt mirror is
about $\pm250$ microns \textbf{that represents approximately 10
arcseconds on the sky, which is enough for the application}.
\textbf{An specific 50Hz algorithm is used for the tip-tilt
mirror. The telescope is tracking at sidereal rate but we do not
autoguide on off-axis guide stars.}

\section{Instrument control}
\subsection{Instrument control electronics.}

SAOLIM is operated by a remote control without direct human
interaction with the instrument. This control electronics were
designed to contain all the subsystems and there were packed into
a single electronic rack located just below the optical bench.
Everything is mounted in a rigid custom-designed aluminium frame.
It contains the following major components: a Pentium IV 3.4Ghz PC
as master hardware controller, a DM Electronics with DC and high
voltage amplifiers, a Tip-Tilt mirror controller, some steeper
motor controllers,  shutter controllers, different power supplies,
and a miscellaneous of support and auxiliary electronic units. The
amount of dissipated heat is negligible with no effect at the
optical bench. However, two fans remove the air inside the
electronic rack. \textbf{They do not cause detectable vibrations.}
A liquid-cooled heat exchanger is installed in this level to
remove the heat produced by the EMCCD inside the optical bench.
This allows us to turn off its fan during the observations at the
telescope, avoiding potential turbulences in the optical path.

The following motorized functions are served by different
electronics sub-systems: a control unit of the SHS shutter, a
linear stage to focus the relay lens onto the SHS, a filter
exchange stage for SHS, and a Tip and Tilt stage of the folding
mirror, E8 in Figure 1, and finally a linear stage to place the
white reference fiber in the optical path.

The PC control is connected via Ethernet through a router
providing direct access from any terminal. A wireless connection
is also available. Such wireless connection is very convenient
when the instrument is controlled next to the telescope for
diagnostics purposes.

\subsection{Instrument control software.}

The system is operated under a Windows XP operating system.
\textbf{Thanks to the very fast processors of modern PCs, all the
procedures can be run under such operating system, without
detectable speed looses. Its use is a new approach and a novelty
which reduces hardware costs considerably.}The control software is
written in Microsoft Visual C++ providing enough computation power
for the reconstruction algorithms and shielding the user from the
detail knowledge of the device parameters.

\subsection{Scientific camera.}

All the tests presented in this article were done using a
non-optimal 1kx1k back-illuminated CCD camera. By the time of
development of the instrument, we did not have access to a more
adequate detector, like a NIR camera, which is the usual
scientific camera in other AO systems. This camera was used only
for testing purposes to check the goodness of the operation of the
reconstruction algorithms. Therefore, no scientifically useful
data were obtained at this stage. All the observations were
performed in the very near infrared regime, using a narrow-band
filter with a central wavelength of 1033nm and a bandwidth of
10nm. \textbf{The transmission was measured and it has no blue
leaks }. At this wavelength range, the camera has still $\sim5\
\%$ quantum efficiency, (QE). For that reason, only bright stars
could be observed. This will not hamper the results presented
here, since our main goal was the test of the design and
construction, not to produce scientifically useful data at this
point.

In a second future stage, a fast frame rate EMCCD camera is
intended to replace the current scientific camera.

\section{Performance of the instrument.}
\subsection{Mechanical stability of the prototype.}
The mechanical design of the instrument is provided by a solid
aluminium cast housing, which keeps all the optical elements in
place. During some bad weather nights, flexure tests at the
telescope were performed. The telescope was pointed to different
positions in hour angle and declination. At each position, the
reference fiber pattern was recorded at the Shack-Hartmann sensor
which computes the resulting Tip and Tilt values of the image. As
a result, flexures of the instrument were determined to be
negligible at any location. Their maximum value was $\sim0.5$
arcsec at 30 degrees of telescope elevation pointing to the south.
In addition, the motorized stages were tested at very low
telescope elevations showing an acceptable behavior, reproducing
the same available positions.
\begin{figure*}
\includegraphics[scale=.60, clip=true]{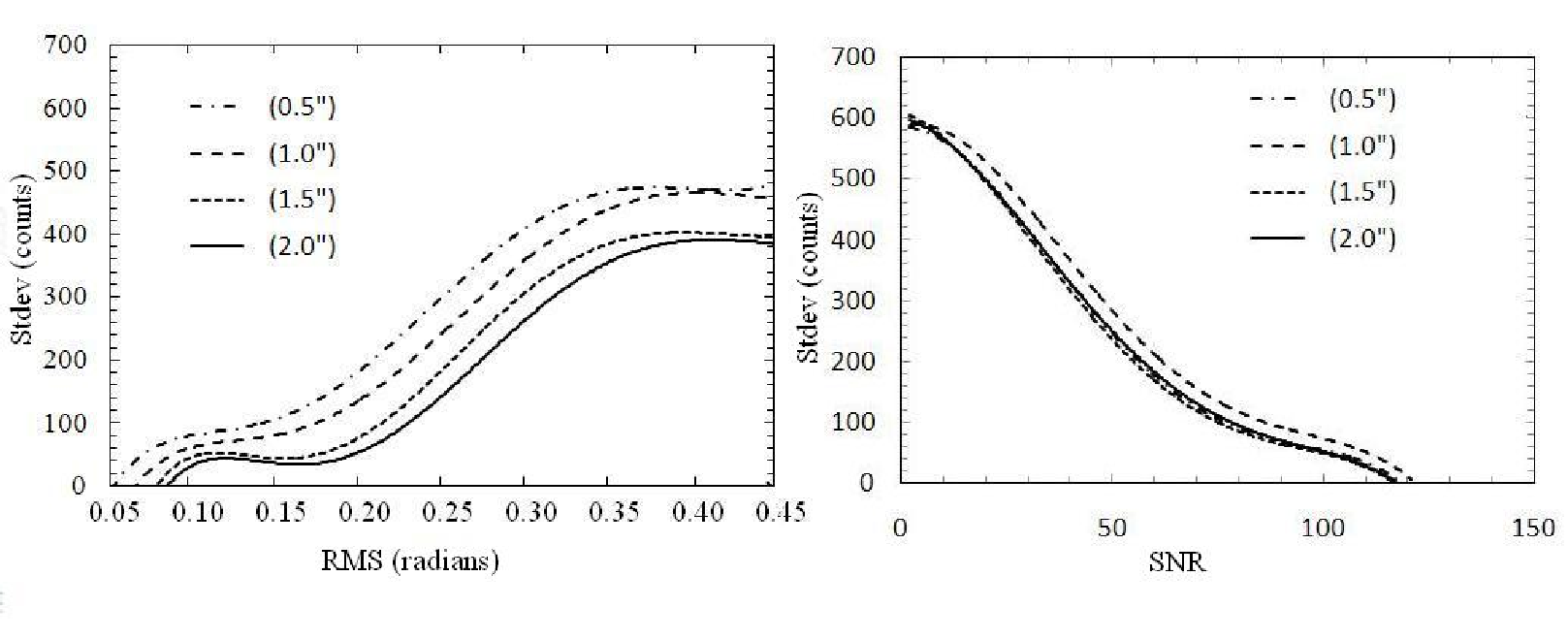}
\caption{Left panel: RMS of the wavefront reconstruction measured
with SHS versus Standard deviation for different seeing
conditions. Units of RMS are in radians. Right panel: SNR per
lenslet channel measured with SHS versus standard deviation for
different seeing conditions. In that case, all the plots are very
similar because the photometric aperture was constant and bigger
than the simulated seeing value.}
\end{figure*}

\subsection{Sensitivity of the WFS.} \label{bozomath}
One innovative aspect in the SHS design was the use of an EMCCD as
a detector. This pioneering camera has had a profound influence on
photon starved imaging applications, as photon counting in
astronomy \citep[e.g.][]{dus2004}. The back-illuminated device
combines photon collection efficiencies of up to 95\% QE with
single photon sensitivity through the virtual elimination of the
readout noise. \textbf{It is likely that some fake detections may
arise. Their effect will be to increase the net rms of the
reconstructed wavefront. For that reason, the final performance of
the instrument has to be measured on real stars.}

During the commissioning of the instrument, in the observing
period of 2008, a set of stars with different brightness were
observed to estimate the overall sensitivity of the instrument.
The results are summarized in Table 2. The photometry was
performed by using a circular aperture of 6 pixels using the
 DAOPHOT package of IRAF. The table shows the star name, V-band
magnitude, total flux in counts and signal-to-noise ratio (SNR).
Each measured parameter in the table represents the average value
over all the sub-apertures. All the images were corrected for dark
current and pixel-to-pixel variations. The table shows the
detection limit of the instrument, which is defined as the maximun
magnitude at which the control software is able to compute a
centroid for every subaperture of the lenslet array and keep
close-loop operation. Such limit is 11.8 mag for SAOLIM. This
detection limit is similar to that one of more complex and
expensive AO systems, like ALFA, mounted at the 3.5m telescope of
the Calar Alto observatory), which was able to use stars as faints
as V$\sim12$ to close the loop \cite{hip2000}.

\begin{table}
\begin{center}
\caption{Sensitivity measurements of SHS. Average seeing FWHM
~1.3"-1.6". Campaign 03-07-2008.\label{tbl-2}}
\begin{tabular}{llllll}
\tableline\tableline Star Name & $m_{v}$ & Total Flux. &
\textbf{SNR{\tablenotemark{a}}} &
EMCCD Setup \\
\tableline
HR7315 & 5.3 & 136785  & \textbf{187.7} & EMGain=100, Freq=135Hz\\
HD171827 & 7.7  & 61428 & \textbf{128.7} &EMGain=210, Freq=103Hz\\
SAO67491 & 8.6  & 10944  & \textbf{52.9} & EMGain=210, Freq=103Hz\\
SAO68044 & 9.7  &  4213 & \textbf{38.5} & EMGain=210, Freq=103Hz\\
PPM82785 & 10.7 & 956  & \textbf{19.1} & EMGain=255, Freq=103Hz\\
GSC2662 & 11.8  &  192 & \textbf{7.5} & EMGain=255, Freq=103Hz\\
\tableline
\end{tabular}
\tablenotetext{a}{\textbf{\cite{rob2003}}}

\end{center}
\end{table}

An empirical relationship between the SNR per sub-aperture
measured by the SHS and the star brightness can be established by
an exponential fitting to both parameters, yielding:
\begin{equation}
SNR = 9538.6_{\pm1.6}e^{-0.4657_{\pm0.015}m_v}
\end{equation}

in that way, the expected SNR can be estimated for any magnitude.
The accuracy of the centroid algorithm during the wavefront
reconstruction is determined by the SNR per sub-aperture.
Therefore, some experiments have to be performed with the aim to
predict the capability of the system to compensate a turbulence
under different seeing conditions and different star magnitudes.

To do so, a reference fiber is fed with a white source and it was
placed exactly at the focal plane of the telescope, simulating a
perfect reference star. By adding white noise to the whole image
in steps of 0.5 counts, the RMS (eq 6) and the SNR  can be
measured as a function of the noise on the different images
recorded by the SHS. Finally, a relation between them can be
established. This procedure was repeated until the SNR dropped to
a low value ($\sim$10), when the uncertainty of the centroid
coordinates was high. Moreover, different seeing conditions were
simulated by convolving the reference fiber pattern image with a
Gaussian function with different widths. In total, 4000
realizations of the RMS and the SNR vs the standard deviation of
the input noise were performed for this simulation. An empirical
relation between the input standard deviation of the simulated
noise, and the output RMS of the reconstructed image and the final
SNR of the detected sub-images was derived for each input seeing
by fitting the simulated data sets with a 5th order polynomial
function, for each pair of parameters. The simulations are shown
in Figure 7.

As a result of the former simulations, the Strehl ratio (at
$\lambda$ = 550nm) can be derived and related to the V magnitude
as shown in Figure 8. Therefore, the faintest star that can be
used as reference by SAOLIM to compensate the turbulence should
have a brightness of at least V$\sim$11.5 with the 28 microlens
array. \textbf{This value was obtained for a natural seeing FWHM
of 1 arcsecond} ($r_{0}$=10cm). The final accuracy of the
correction will depend strongly on the input natural seeing
conditions.

\begin{figure}
\includegraphics[scale=.50, clip=true]{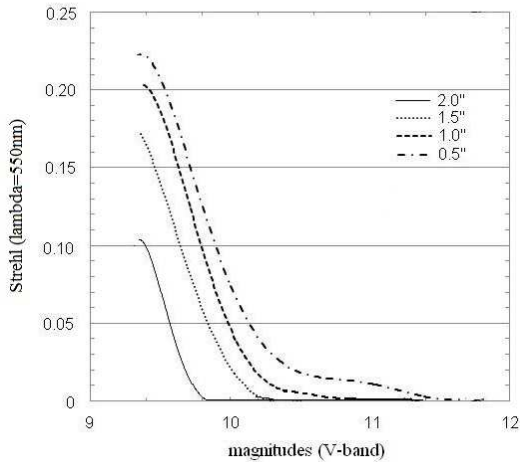}
\caption{\textbf{Simulated} relation between the Strehl ratio and
V magnitude measured by SHS for different seeing conditions.}
\end{figure}

\section{Results on the sky.} \label{bozomath}

Four technical campaigns were carried out to test the performance
of the instrument on real stars at the 2.2m telescope of the Calar
Alto observatory. The goals for these observing runs are
summarized in the Table 3.
\begin{table}
\begin{center}
\caption{Technical runs for the commissioning of SAOLIM during
2007 and 2008 in the observatory of Calar Alto
observatory.\label{tbl-2}}
\begin{tabular}{lll}
\tableline\tableline Date  & Goals\\
\tableline \tableline\\
(2-5)-07-2007 & Determination of the pixel scale.\\
 & Flexures. Wavefront sensitivity.\\
(28-30)-09-2007 & First image obtained on the\\
 &  scientific camera of a real\\
  & star with the Tip Tilt compensation. \\
(11-14)-05-2008 & First images obtained on the\\
 & scientific camera of a real\\
  & star with the high order compensation\\
  & (under very bad seeing conditions,\\
  & performance from 2.5" to 1.2"). \\
16-08-2008 & Evaluation of the performance\\
 & of the high order loop on the\\
 & scientific camera under average\\
 & seeing condition. \\

\tableline
\end{tabular}
\end{center}
\end{table}
\subsection{Static aberrations of the telescope.}


The static aberrations of the 2.2m telescope are well known (U.
Thiele private communications). They were analyzed previously by
using the intra-extra focal images technique \cite{van2002}. This
method provides a set of Zernike coefficients that characterizes
those aberrations. To check the reliability of the reconstruction
algorithm adopted in SAOLIM, a routine was implemented in the
control software to determine them. For doing so, the routine
computes the wavefront coefficients of 10000 images. Under good
seeing conditions the mean values of the coefficients represent
the static aberrations of the telescope. Table 4 lists the
coefficients of the static aberration of the telescope measured by
the method described before. There is a very good agreement
between both estimations showing the capability of the instrument
to analyze correctly the turbulence aberrations.

\begin{table}
\begin{center}
\caption{Comparison between static coefficients of the 2.2m
telescope at Calar alto obtained with the intra-extra focal images
and SAOLIM wavefront sensor.\label{tbl-2}}
\begin{tabular}{lrr}
\tableline\tableline Aberration  & RMS CAHA(nm) & RMS SAOLIM
(nm)\\
\tableline \tableline\\
Astigmatism (sin) & -7.54  & -6.5$\pm$1.0\\
Astigmatism (cos) & 21.2   & 22.2$\pm$1.0\\
Coma (sin)        & 5.36   & 4.3$\pm$0.6\\
Coma (cos)        & 8.42   & 10.1$\pm$0.5\\
Trifoil (sin)     & 4.81   & 3.7$\pm$0.6\\
Trifoil (cos)     & 5.21   & 5.8$\pm$0.5\\
Spherical         & 13.1   & 12.9$\pm$0.3\\
Quad astig (sin)  &-0.50   &-0.6$\pm$0.3\\
Quad astig (cos)  &-2.02   &-2.5$\pm$0.3\\
\tableline
\end{tabular}
\end{center}
\end{table}

\subsection{The Tip-Tilt algorithm on a real star.}

In order to test the Tip-Tilt compensation algorithm a real star
(HR7331) was observed under average seeing FWHM of 1.3". The star
was not centered on the detector on purpose. The drift of the
centroids imaged by the lenslet array with respect to a reference
image produced by a fiber placed at the focal plane of the
telescope, are computed by the wavefront sensor. With a proper
calibration, the Tip-Tilt mirror is commanded in such a way that
the drift is minimized in both axes.

Figure 9 shows the variations of the centroid of the star. They
are minimized when the compensation is imposed in such a way that
the standard deviation of the measurements decreases from 0.27 to
0.16 pixels. Besides this, the star is brought to the center of
the SHS by the system. An improvement of 25\% in terms of the FHWM
and of 40\% in terms of peak intensity is measured on the
scientific camera as shown in Figure 10.

\begin{figure}
\includegraphics[angle=270,scale=.30]{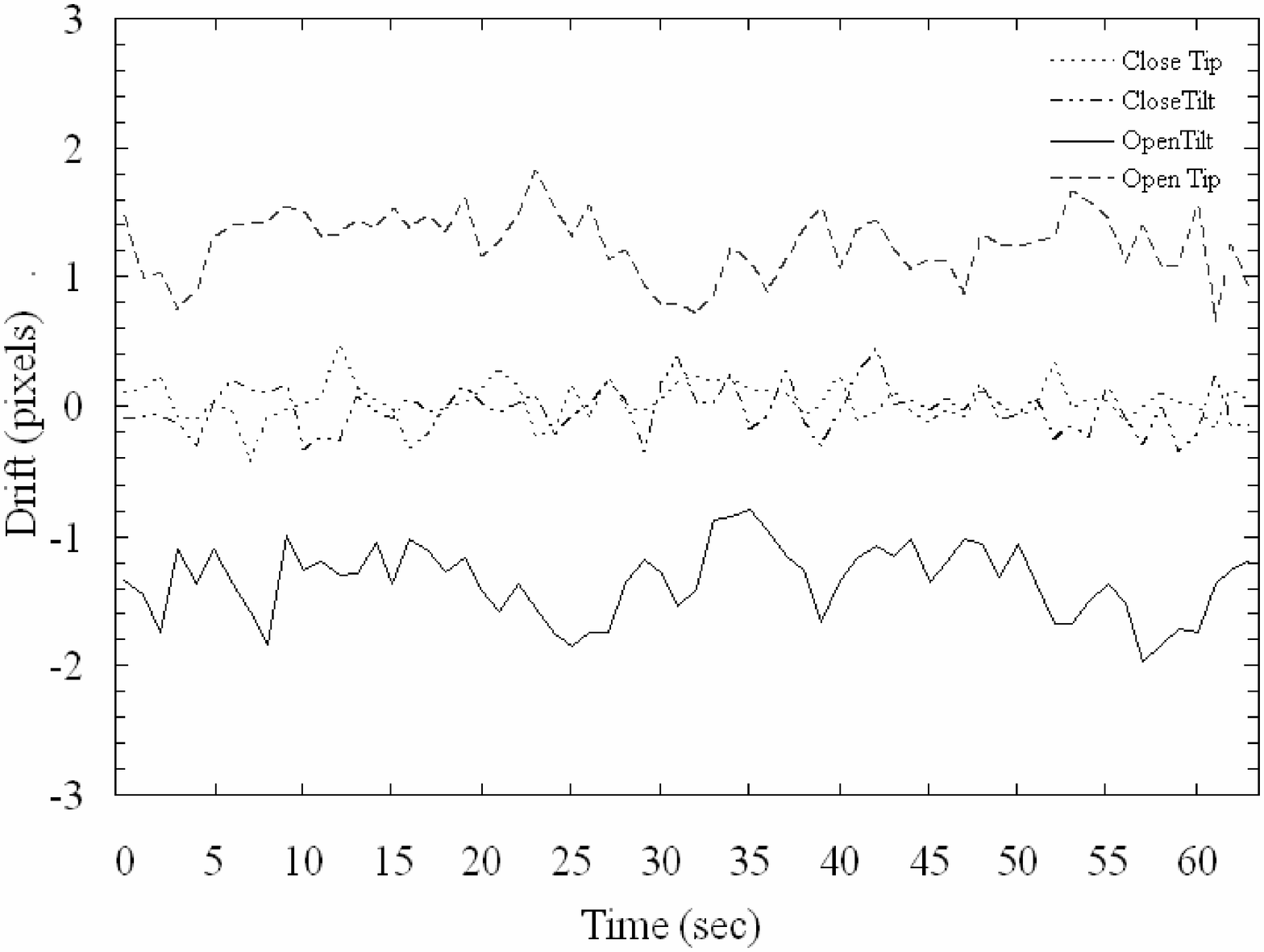}
\caption{Centroid distances of the SH star pattern with respect to
the SH reference fiber when a Tip-Tilt compensation is applied.}
\end{figure}
\begin{figure}
\includegraphics[scale=.30, clip=true]{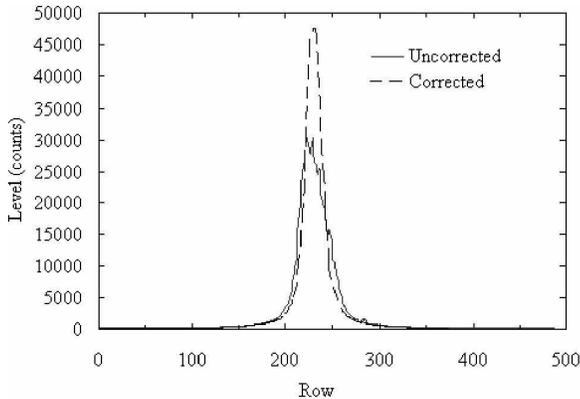}
\caption{Horizontal intensity cut of HR7331 image on the
scientific camera. An improvement of 25\% in FWHM and 40\% in peak
intensity is observed when the Tip-Tilt compensation is applied.
The natural seeing FWHM was about 1.3". Corrected seeing FWHM was
1.02".}
\end{figure}

\subsection{High order closed-loop results.}

The high order algorithm (HO) was also tested with real
observations on the sky. The test comprises the measurements of
the \emph{RMS} of the wavefront (eq.6) during the open-loop and
the closed-loop procedures. In the ideal case, this magnitude
should be zero when the wavefront compensation is active which
will mean that the wavefront is flat. In the closed-loop operation
when the SHS only measures the residual image motion after each
correction, the measured values are smaller, than in open loop
operation.

Figure 11 shows the \emph{RMS} measured by the wavefront sensor
during the open-loop procedure, the Tip-Tilt and the high order
compensation procedures as a test to check the smooth performance
of the reconstruction algorithms. Again, the units are radians.
Clear differences between the three processes are appreciated. A
natural guide star of V $\sim5.2$ magnitude was used for this
experiment, with a SHS frame rate of 420Hz and collecting a total
number of 21000 \textbf{counts} for each process. The average
seeing FWHM conditions were around 1.4". An improvement of a
factor $\sim2-3$ is observed in terms of measured \emph{RMS} by
the Shack-Hartman sensor when only the Tip-Tilt compensation (TT)
is applied. When the high order algorithm (HO) is added, a
decrease of the RMS by a factor 15 is seen. On the other hand, a
decrease in the median values of every mode's coefficient is
observed. The decrease for the modes from 2-10 is listed in Table
5.

\begin{figure}
\includegraphics[scale=.42, clip=true]{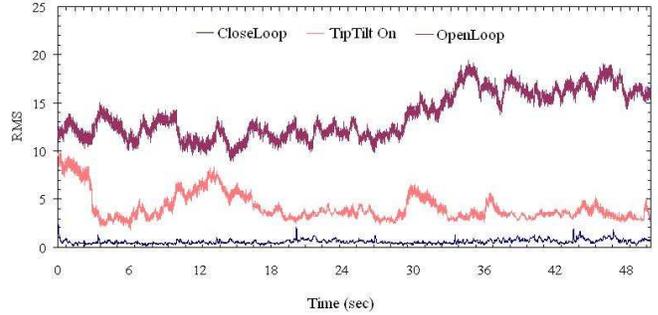}
\caption{\emph{RMS} of the wavefront during Open-Loop, Close-TT
and Closed-Loop for a real natural star. Units are given in
radians for \emph{$D/r_{0}$} = 22.}
\end{figure}

\begin{table}
\begin{center}
\caption{Coefficients of the modes 2-10 with and without HO
compensation. Values are in microns.\label{tbl-2}}
\begin{tabular}{lll}
\tableline
mode & Open-loop & HO+TT\\
\tableline
2 & 2.59 & 0.015\\
3 & 0.96 &0.002 \\
4 & -0.45 & -0.011\\
5 & -0.41 & -0.002\\
6 & 0.13 & -0.006\\
7 & -1.33 & -0.14\\
8 & -0.05 & 0.11\\
9 & -0.17 & -0.08\\
10 & 0.22 & 0.21\\
\tableline
Average & 1.04 & 0.09\\
\tableline
\end{tabular}
\end{center}
\end{table}

Any AO system may have internal inconsistencies: i.e., the system
considers that the correction is adequate, minimizing the RMS, but
the wavefront is not correctly compensated. For instance, this may
happen if the wavefront sensor has a systematic error. Then the
influence functions for the membrane would be obtained with that
bias, and the global failure of the system would not be noticeable
only from the analysis of the RMS.

To be completely sure of the performance of our AO system,
simultaneous images were taken with the scientific camera. Figures
12-14 show some examples of a real-time closed-loop aberration
compensation using a natural guide star. These preliminary results
were obtained during the commissioning of the instrument at the
2.2m Calar Alto telescope. An observing log of the observations
carried out can be found in table 6.

The images are clearly improved when applying the closed loop
wavefront compensation even under quite poor seeing conditions.
Figure 12 shows a 4.6 magnitude star (SAO88071) observed under
very bad turbulence conditions, with a natural seeing FWHM of
about 2.5". The data were obtained during the campaign of May
2008. Under such bad seeing, 30\% of the actuators reached their
maximum values and therefore only an improvement of a factor of 2
in terms of FWHM could be achieved. The central wavelength of the
observations was again the same as before. The frame rate of the
reconstruction was 420Hz and 14 modes were taken into account with
the KS28 lenslet array. Under these poor observational conditions,
observing techniques like lucky imaging are completely useless.
However SAOLIM was able to reduce the FWHM of the output image to
half of its input value and to increase the peak intensity by a
factor $\sim6$. Figure 13 illustrates that correction. Since the
lucky imaging technique is feasible with an input seeing FWHM of
$\sim$1" \cite{hor2008}, this experiment demonstrates how our
instrument can improve the performance of a lucky imaging device.

\begin{table*}
\begin{center}
\caption{Observing log for some of the results obtained with the
scientific camera of SAOLIM during the observing run at the
telescope. A narrow band filter (1033/10nm) was used during all
the observations.\label{tbl-2}}
\begin{tabular}{llllllll}
\tableline\tableline Date & UT & Objectname & V mag& Exp. Time & Loop
parameters & FWHM$_{\rm Uncorrected}$ & FWHM$_{\rm Corrected}$ \\
\tableline

2008-05-13 & 02:00 & SAO88071 & 4.6 & 10 sec & 420Hz 14 modes & 2.5" & 1.2''\\
2008-09-16 & 19:46 & SAO103052 & 6.2 & 10 sec & 420Hz 16 modes & 1.2'' & 0.37'' \\
2008-09-17 & 02:40 & WDS01095+4795 & 4.59 and 5.61 & 10 sec & 300Hz 16 modes&
1.1'' & 0.32''\\

\tableline
\end{tabular}
\end{center}
\end{table*}

During the campaign of September 2008, the system was tested under
better seeing conditions. In the left panel of Figure 14, an image
of the double star WDS01095+4795 (with a visual magnitude of 4.59
and 5.61) is shown, taken with SAOLIM, without applying any
corrections. The natural seeing FWHM was 1.1" during the
observations. The separation of the star is 0.4" hence it was
unresolved and appeared as a single spot in the image. The right
panel shows the same object observed with the real-time
closed-loop active. The corrected FWHM was 0.32" and as a result
of this, the double star is clearly resolved. The frame rate of
the loop was 300 Hz.

This result is quite promising if the prototype is attached to a
lucky imaging camera. When combined with an AO system, Lucky
Imaging selects the periods when the turbulence that the adaptive
optics system must correct is reduced. In these periods, lasting a
small fraction of a second, the correction given by the AO system
is sufficient to give excellent resolution with visible light. The
Lucky Imaging system sums the images taken during the excellent
periods to produce a final image with much higher resolution than
is possible with a conventional long-exposure AO camera
\cite{law2006}.

\begin{figure}
\includegraphics[scale=.30, clip=true]{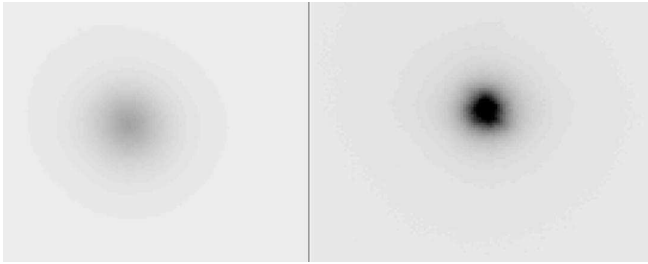}
\caption{Left panel: Image of the star SAO88071 taken with the
scientific camera of SAOLIM during openloop. Both images are a
subwindow of 4x4arcsec. Natural seeing FWHM was 2.5 arcseconds.
Right panel: Same star under same seeing conditions during closed
loop procedure. Corrected seeing FWHM was 1.2 arcseconds. The
images were taken with a narrow filter with a central wavelength
of 1033nm and 10nm of bandwidth from the Calar Alto 2.2m
telescope. The reference star has a 4.66 V-band magnitude.}
\end{figure}
\begin{figure}\includegraphics[scale=.40, clip=true]{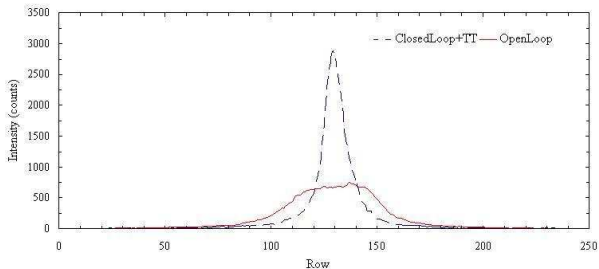}
\caption{Intensity profiles of the open and closed loop images
shown in Fig 12.}
\end{figure}

\begin{figure}
\includegraphics[scale=.60, clip=true]{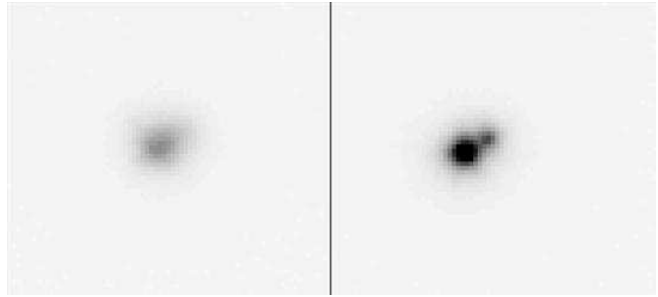}
\caption{Images of the double star WDS01095+4795 of 4.59 and 5.61
magnitudes respectively taken with the scientific camera of
SAOLIM. Both images are a subwindow of 4 arcsec$^{2}$. Left panel:
Open loop. Right panel: Closed loop. The loop frequency was 300Hz
and 16 modes were corrected. The double star has a separation of
0.4arcsec and it is perfectly resolved. The central wavelength of
the observations was 1033/10nm. Both images are displayed with the
same cut levels. Natural seeing FWHM was 1.1arcsec and the
corrected one was 0.32arcsec. The peak intensity rose up from 2300
to 10200 counts. }
\end{figure}

\section{Conclusions}
A low cost adaptive optic system was developed for astronomy and
tested. It uses a 39-actuator membrane deformable mirror of 20mm
diameter, a fast frame rate EMCCD as detector of the wavefront
sensor and a low cost Tip-Tilt mirror. The whole prototype is
running in a single PC, resulting in a compact module
easy-to-install and transport with a total weight of only 70kg.
Experimental simulations were carried out to determine the
sensitivity of the wfs and to predict the limit magnitude of the
star to be used. This limit was found to be 11.5 magnitudes. The
system was made up of entirely commercial  hardware components
with a total cost of about 35000 euros. We did not count here
\textbf{the manpower costs}.

A powerful method was adopted to evaluate the capabilities of a
membrane deformable mirror to produce and correct different
aberrations within the range of interest in astronomy. This is a
general approach, suitable for use in every system whose control
was based on a previous knowledge of its influence functions.
Under the assumption of linearity, the proposed iterative
algorithm works with the required precision, making it appropriate
to use in real-time applications. Karhunen-Loeve's polynomial or
any arbitrary surface could be reproduced when this procedure is
systematically applied, taking into account the available range of
voltages of the mirror. A real-time (up to 420Hz) closed-loop
algorithm has been incorporated to the device for the compensation
of atmospheric turbulence.

The system has been tested on real stars. The best images obtained
had a FWHM of 0.32 arcsec from an input natural seeing FWHM of 1.1
arcsec. Although the system does not achieve diffraction limited
images, the FWHM is improved by a factor of 4. Under quite poor
seeing conditions (FWHM=2.5") 30\% of the DM actuators were
saturated, but an improvement of a factor of 2 was measured in the
FWHM of the images. The Tip-Tilt mirror showed a good behavior too
reducing the FWHM from 1.3" to 1.02".

The use of this AO system attached to a Lucky imaging system could
enhance the spatial resolution of the input images and the
fraction of useful images, and therefore the performance of such a
system will be increased. Both techniques will be tested in the
future.

\acknowledgments

We are grateful to Plan nacional I+D+I de Astronomia y Astrofisica
AYA2005-07808-C03-01 and AYA2008-06202-C03-01 that supported this
project. Also FEDER funds are acknowledged. The authors are
grateful to the Plan Andaluz de Investigacion, Desarrollo e
Innovacion, who granted this study under the projects:
P08-FWM-04319 and FQM360. They would also thanks the ICTS-2009-32
program of the spanish Ministerio de Ciencia e Innovación. The
paper is based on observations collected at the Centro Astronómico
Hispano Alemán (CAHA) at Calar Alto, operated jointly by the
Max-Planck Institut für Astronomie and the Instituto de
Astrofísica de Andalucía (CSIC). We are grateful to Luzma Maria
Montoya and Santos Pedraz for their help during the observations.
We would like to thank the anonymous referee for the useful
comments.

SFS would like to thanks the {\it Fundaci\'on Agencia Aragonesa para la
Investigaci\'on y el Desarrollo} (ARAID), for the financial support during the
last year.

\clearpage



\clearpage








\clearpage


\clearpage


\clearpage


\clearpage




\end{document}